# IMPACT OF THE DOMAIN STRUCTURE IN FERROELECTRIC SUBSTRATE ON GRAPHENE CONDUCTANCE (AUTHOR REVIEW)

Maksym V. Strikha[1,2*], Anatolii I. Kurchak[1†], and Anna N. Morozovska[3‡],

[1] *V.Lashkariov Institute of Semiconductor Physics, National Academy of Sciences of Ukraine,*

*pr. Nauky 41, 03028 Kyiv, Ukraine*

[2] *Taras Shevchenko Kyiv National University, Radiophysical Faculty*

*pr. Akademika Hlushkova 4g, 03022 Kyiv, Ukraine*

[3] *Institute of Physics, National Academy of Sciences of Ukraine,*

*pr. Nauky 46, 03028 Kyiv, Ukraine*

## Abstract

Review is devoted to the recent theoretical studies of the impact of domain structure of ferroelectric substrate on graphene conductance. An analytical description of the hysteresis memory effect in a field effect transistor based on graphene-on-ferroelectric, taking into account absorbed dipole layers on the free surface of graphene and localized states on its interfaces is considered. The aspects of the recently developed theory of p-n junctions conductivity in a graphene channel on a ferroelectric substrate, which are created by a 180-degree ferroelectric domain structure, are analyzed, and cases of different current regimes from ballistic to diffusion one are considered. The influence of size effects in such systems and the possibility of using the results for improving the characteristics of field effect transistors with a graphene channel, non-volatile ferroelectric memory cells with random access, sensors, as well as for miniaturization of various devices of functional nanoelectronics are discussed.

---

[*] author 1, e-mail: maksym.strikha@gmail.com
[†] author 2, e-mail: teoretyk0706@gmail.com
[‡] author 3, e-mail: anna.n.morozovska@gmail.com

# ВПЛИВ ДОМЕННОЇ СТРУКТУРИ СЕГНЕТОЕЛЕКТРИЧНОЇ ПІДКЛАДКИ НА ПРОВІДНІСТЬ ГРАФЕНУ (АВТОРСЬКИЙ ОГЛЯД)

Максим В. Стріха[1,2], Анатолій І. Курчак[1], Ганна М. Морозовська[3],

1 Інститут фізики напівпровідників ім. В.Лашкарьова Національної академії наук України, пр. Науки 41, 03028 Київ, Україна

2 Київський національний університет імені Тараса Шевченка, радіофізичний факультет, пр. Академіка Глушкова 4г, 03022 Київ, Україна

[3] Інститут фізики НАН України, пр. Науки 46, 03028 Київ, Україна

**Анотація**

Огляд присвячено останнім теоретичним дослідженням впливу доменної структури сегнетоелектричної підкладки на провідність графенового каналу. Розглянутий аналітичний опис ефектів пам'яті гістерезисного типу у польовому транзисторі на основі графена-на-сегнетоелектрику, з урахуванням адсорбованих дипольних шарів на вільній поверхні графену і локалізованих станів на його інтерфейсах. Аналізуються аспекти нещодавно розвинутої теорії провідності p-n переходів у графеновому каналі на сегнетоелектричній підкладці, які створені 180-градусною сегнетоелектричною доменною структурою, причому розглянуті випадки різних режимів струму, від балістичного до дифузійного. Обговорюється вплив розмірних ефектів у таких системах та можливість використання результатів для вдосконалення характеристик польових транзисторів з графеновим каналом, комірок енергонезалежної сегнетоелектричної пам'яті з довільним доступом, сенсорів, а також для мініатюризації різних пристроїв функціональної наноелектроніки

**Ключові слова:** графен-на-сегнетоелектрику, доменна структура, провідність, польовий транзистор

# I. INTRODUCTION

Experimental and theoretical studies of the remarkable electromechanical, electrophysical and transport properties of graphene remain on the top of researchers' attention since graphene discovery [1, 2] till nowadays (see e.g. [3, 4, 5]). A promising and quite feasible way towards understanding and control of the graphene-based devices (as well as the devices utilizing other 2D semiconductors) is to use "smart" substrates with additional (electromechanical, polar and/or magnetic) degrees of functionality. For instance, a graphene on a ferroelectric substrate [6, 7, 8, 9, 10], whose spontaneous polarization and domain structure can be controlled by an external electric field [11, 12], can be proposed as such "smart" system [13].

The author review is organized as following. Section II analyses our recent theoretical works devoted to the changes of graphene channel conductivity caused by p-n junctions induced by the existence of single domain wall in a ferroelectric substrate, discuss the different regimes of current and rectification effects. Section III is devoted to the discussion and analyses of the theory of hysteretic phenomena in graphene-on-ferroelectric. Section IV analyses the dynamics of p-n junctions in graphene channel induced by the motion of ferroelectric domain walls. Section V discusses the possibilities of graphene separation and stretching induced by piezoelectric effect of ferroelectric domains. Section VI presents a brief summary of the results.

# II. CONDUCTIVITY OF GRAPHENE CHANNEL WITH P-N JUNCTION ON FERROELECTRIC DOMAIN WALL

The presence of a domain structure in a ferroelectric substrate can lead to the formation of p-n junctions in graphene [11, 12], which are located above the domain walls in a ferroelectric substrate [14, 15, 16, 17, 18]. Note that the unique properties of the p-n junction in graphene have been realized much earlier by multiple gates doping of graphene channel by electrons or holes, respectively [19, 20, 21]. Then they have been studied theoretically [22, 23] and experimentally [24, 25, 26]. Hinnefeld et al [11] and Baeumer et al [12] created a p-n junction in graphene using ferroelectric substrates by imposing a graphene sheet on a 180°-ferroelectric domain wall (**FDW**). Due to the charge separation by an electric field of a FDW – surface junction, p-n junctions occur in graphene [27, 28].

Note that the elastic strain can change the band structure of graphene (e.g. via the deformation potential) and open the band gap [4, 5, 29, 30, 31]. We have shown [18] that the piezoelectric displacement of the ferroelectric domain surfaces can lead to the stretching and separation of graphene areas at the steps between elongated and contracted domains. Graphene separation at FDWs induced by piezo-effect can cause the increase of graphene channel conductance. Below we briefly review the results of Refs.[14-18] and discuss their correlation with available experimental and other theoretical works.

Semi-quantum and semi-phenomenological analytical models have been developed for different types of carrier transport (ballistic, diffusive, etc) in a single-layer graphene channel at 180°-FDW [14, 15].

Specifically, the influence of a ferroelectric domain wall on the ballistic conductance of a graphene channel has been studied in the Wentzel-Kramers-Brillouin (**WKB**) approximation in Ref.[14]. Here we considered the graphene channel separated from a ferroelectric layer with the FDW structure by an ultra-thin dielectric layer (a physical gap or a ferroelectric dead layer). The ferroelectric layer makes an ideal electric contact with the gate electrode [see **Fig.2.1(a)**].

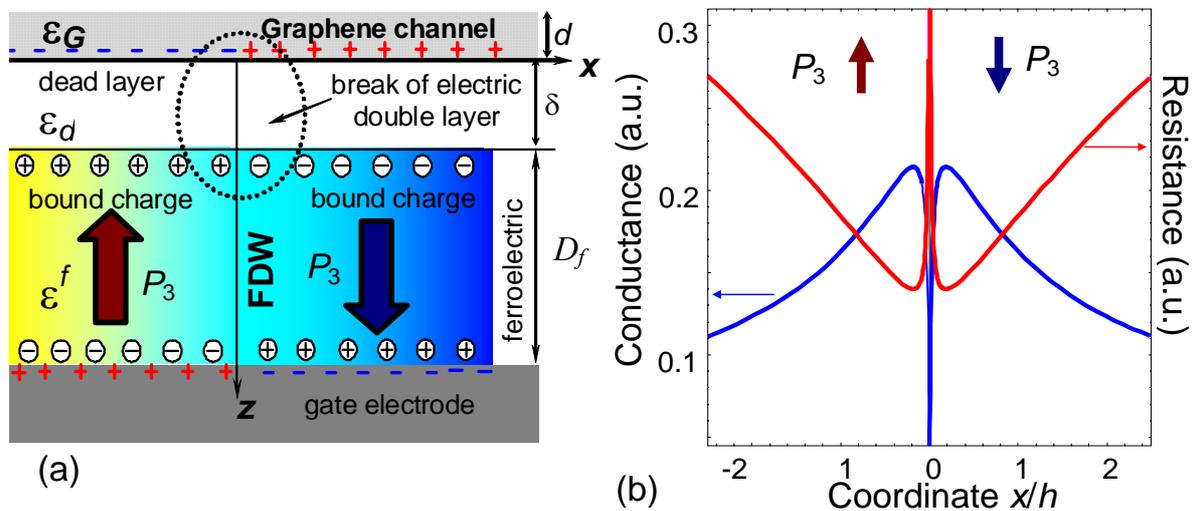

**Figure 2.1. (a)** 180°-FDW structure near the ferroelectric surface in the graphene-on-ferrroelectric. The discontinuity of the electric double layer consisting of screening and bound charges creates a depolarization electric field that penetrates into the gap. **(b)** Graphene conductance (left vertical axis) and resistance (right vertical axis) $x$-distributions caused by the FDW located at $x=0$ in the graphene-on-ferrroelectric heterostructure. (Reproduced from [A. N. Morozovska, E. A. Eliseev, and M. V. Strikha. Applied Physics Letters **108**, 232902 (2016)], with the permission of AIP Publishing).

A pronounced broadening of the FDW near the surface appears [14, 32] in order to decrease depolarization electric field produced by the uncompensated polarization bound charges localized in a thin sub-surface layer of ferroelectric. Despite the broadening, the stray electric field is still strong enough to induce the p-n junction in the graphene channel. A potential barrier for electrons and holes emerges in the p-n junction, when a positive (negative) voltage is applied to the left (right) contact of the channel. The probability $w$ for an electron in the *n*-region with the wave vector $k$ directed at the angle $\vartheta$ with respect to the $x$ axis to pass into the *p*-region can be calculated using the scheme presented in Ref.[33]. With an allowance for the linear graphene band spectrum $E(k_x, k_y) = v_F \hbar \sqrt{k_x^2 + k_y^2}$, at the junction centre, $x=0$, the electron's kinetic energy equals $v_F \hbar \sqrt{k_x^2 + k_y^2}$, where the momentum $y$-component $k_y = k_F \sin \vartheta$ is conserved ($v_F \approx 10^6$ m/s is so called Fermi velocity of the electron, it is determined by the energy of σ-bonds between carbon atoms in graphene plane, Ref.[3]). Therefore, the $x$ component is determined by the expression $k_x(x) = \sqrt{(e\varphi(x,0)/v_F \hbar)^2 - (k_F \sin \vartheta)^2}$. The classically allowed region of electron motion is determined by the inequality $e\varphi(x,0) > v_F \hbar k_y$, which means that the electron cannot overpass the turning point at the distance $l_x$ from the centre of the junction located at the FDW.

For small angles $|\vartheta| \ll 1$, the probability $w$ can be estimated quasi-classically in the WKB approximation as $w \approx e^{-2S/\hbar}$, where $S = i\hbar \int_{-l_x}^{l_x} k_x(x)dx$. The conductance *per p-n junction unit width* has been estimated as [14]:

$$G_{pn}^{ball} \approx \frac{e^2 k_F}{2\pi^2 \hbar} \sqrt{\frac{\pi}{P_x}}. \tag{2.1}$$

In Eq.(2.1) $k_F = \frac{\pi^2 \varepsilon_0 \varepsilon_{33}^f}{e^2} \frac{\hbar v_F Q}{\delta^* \ln 3}$ is the Fermi wave vector, $Q$ is the dimensionless factor $Q = \frac{e(P_S/\varepsilon_0)}{\varepsilon_{33}^f + \gamma \varepsilon_d} \frac{2\gamma \delta^* \ln 3}{\pi \hbar k_F v_F}$ and $P_x \approx \pi^3 \varepsilon_0 \varepsilon_{33}^f \hbar v_F / e^2$ is the probability factor. The spontaneous polarization is $P_S$, $e$ is an electron charge, $\varepsilon_0$ is the universal dielectric constant, the relative permittivity of the dielectric layer $\varepsilon_d$ is equal either to the background permittivity of ferroelectric for the dead layer or to unity for the physical gap; $\gamma = \sqrt{\varepsilon_{33}^f/\varepsilon_{11}^f}$ is the anisotropy factor of ferroelectric, $\varepsilon_{ii}^f$ are the components of relative permittivity of ferroelectric substrate, ($\varepsilon_{33}^f$ is the permittivity of ferroelectric along z direction, that is normal to the graphene plane, $\varepsilon_{11}^f$ is the permittivity of ferroelectric along channel x-direction); $\delta^*$ is the sum of the thickness δ of dielectric layer and effective screening length of graphene [14, 34].

Since the wave vector $k_F \equiv \sqrt{\pi n_{2D}}$, (2.1) can be rewritten as [14]:

$$G_{pn}^{ball} \cong \frac{e^2}{\pi \hbar} \sqrt{\frac{\alpha}{\varepsilon_{33}^f} \frac{c}{\pi v_F} n_{2D}} \tag{2.2}$$

Here, $\alpha = e^2/4\pi\varepsilon_0(\hbar c)$ is the fine-structure constant, and $c$ the light velocity in vacuum. With an accuracy of a dimensionless factor $\sqrt{\alpha c/(\varepsilon_{33}^f \pi v_F)} \sim 0.1$, Eq.(2.2) coincides with the bookish expression for the ballistic conductivity of graphene per unit channel width, $\sigma = 2e^2\sqrt{n_{2D}}/\hbar$ [3, 35]. However, the Eq.(2.2) differs from Eq.(2) in [33], because in for graphene-on-ferroelectric the concentration of 2D-carriers in graphene $n_{2D}$ is governed by the spontaneous polarization of ferroelectric, $n_{2D}(x) \approx \dfrac{2\gamma\varepsilon_{33}^f(P_S/e)}{\varepsilon_{33}^f + \gamma\varepsilon_d}$. The nominal resistance $R_{np}$ is inversely proportional to the conductance (2.2), i.e. $R_{np} = 1/(WG_{pn}^{ball})$, where $W$ is the p-n junction width.

The conductivity and resistivity redistributions caused by the FDW are shown in **Figure 2.1(b)**. As one can see, the width of the p-n junction region is about $2h$. Since $\sigma_0 = e^2/(2\pi\hbar)$ is a quantum of conductance, the ratio $WG_{pn}^{ball}/\sigma_0$ gives us a notion about the number of conductivity modes. This result coincides by order with that calculated in [36].

The estimation of the concentration caused by ferroelectric dipoles leads to values of about $10^{19}$ m$^{-2}$ that is two orders of magnitude higher than for the highest one possible for gated graphene on the SiO$_2$ substrate (this limit is determined by the dielectric breakdown field). Therefore, the graphene p-n junction at the FDW would be characterized by rather a high ballistic conductivity and a low resistivity. However, presence of the factor $1/(\varepsilon_{33}^f)^{1/2}$ in (2.2), where $\varepsilon_{33}^f$ can be rather high for the ferroelectric substrate, for the conduction of p-n junction for the direct polarity of voltage on S and D electrodes (electron is moving through the barrier of p-n-junction), on one hand, and the absence of this factor for the opposite polarity (electron doesn't feel the barrier, while moving in the opposite direction), on the other, causes that graphene p-n junction at the ferroelectric domain would be an excellent rectifier with a conductivity ratio of about 10 between the direct and reverse polarities of applied voltage.

However, the majority of realistic graphene devices are described by the mean free path of electrons $\lambda \sim (50\text{-}250)$ nm and so are operating in a diffusive regime [15]. Since the electron mean free path $\lambda$ in graphene channel is usually much longer than the intrinsic thickness $w$ of the uncharged domain wall in proper ferroelectric, that is about $1 - 10$ nm [37], one can divide the graphene channel with the length $L$ between the source and the drain into 3 sections: one, containing p-n junction itself, with the length $\lambda$, and the two others, on both sides from the junction, with total length $L - \lambda$ (see **Fig.2.2**).

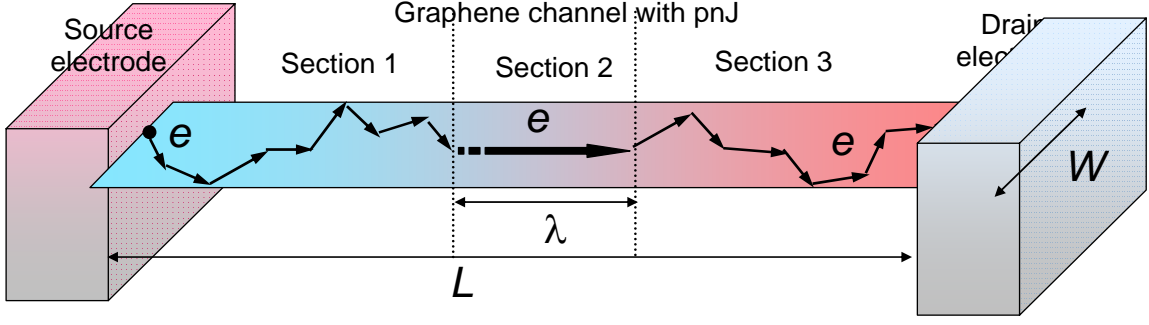

**Figure 2.2.** Formal division of graphene channel with p-n junction (pnJ) into three sections. The current in Sections 1, 3 is diffusive, while in Section 2 it is ballistic. (Reproduced from [M.V. Strikha, A. N. Morozovska. J. Appl. Phys. **120**, 214101 (2016)], with the permission of AIP Publishing).

The total conduction of the sample is governed by evident expression [15]

$$\frac{1}{G^{total}} = \frac{1}{W}\left(\frac{1}{G^{diff}} + \frac{1}{G^{ball}}\right), \tag{2.3}$$

As anticipated the conduction is proportional to the graphene channel width $W$ and the bookish relationship $G^{diff} = \frac{\lambda}{L+\lambda} G^{ball}$ is valid (see e.g. Refs. [40, 51]). The change of voltage polarity on contacts does not change the conduction of the sections 1 and 3. On the contrary, the unit length conduction of the section 2 is described in the cases by different expressions, $G^{ball} = \frac{2e^2}{\hbar \pi^{3/2}}\sqrt{n_{2D}}$ and $G^{ball}_{pn} \cong \frac{e^2}{\pi \hbar}\sqrt{\frac{\alpha}{\varepsilon^f_{33}}\frac{c}{\pi v_F}} n_{2D}$ (see Eq.(2.2) and Ref.[15]), and the permittivity $\varepsilon^f_{33}$ of ferroelectric along the direction normal to graphene plane can be rather high (~ 500 for PZT or even much higher ~5000 for relaxor ferroelectric). Thus the ratio of conductions for different polarities is:

$$\frac{G^{total}_+}{G^{total}_-} = \frac{\beta(L+\lambda)}{\beta(L+\lambda)+\lambda} \tag{2.4}$$

The parameter $\beta$ depends on the dielectric permittivity, $\beta(\varepsilon^f_{33}) = \sqrt{\pi \alpha c/(4\varepsilon^f_{33} v_F)}$ and the electron mean free path $\lambda$ depends on the concentration $n_{2D}$, and so on ferroelectric polarization, since $n_{2D} \sim \pi(P_S/e)$ [15, 34]. When the scattering of electrons in graphene channel at charged impurities in ferroelectric is dominant (which is a most common case for the real graphene operational device, see e.g. [3]), $\lambda(n_{2D}) \sim \sqrt{n_{2D}}$. For the case of the short ballistic channel, $L \ll \lambda$, $\frac{1}{G^{total}} \approx \frac{1}{G^{ball}}$. In the opposite limit of the long diffusive channel, $L \gg \lambda$, $\frac{G^{total}_+}{G^{total}_-} = 1$ and so the rectifying properties of p-n junctions vanish.

The ratio $G_+^{total}/G_-^{total}$ as a function of ratio $L/\lambda$ calculated for different ferroelectric substrates (relaxor, Pb(Zr,Ti)O$_3$, BaTiO$_3$, LiTaO$_3$ and LiNbO$_3$) is presented in **Fig. 1.3**. As one can see from **Fig. 1.3**, small values of $G_+^{total}/G_-^{total}$ can be obtained for the ratios of $L/\lambda$ of unity order or smaller and ferroelectric substrates with high permittivity. The expected result can be realized for the channels of sub-micron length in the case of the comparative values of the electron mean free path. The situation corresponds to the current regime transient from diffusive to ballistic one; mention, that the distance between the parallel FDW is generally much longer, which enables to fabricate a micron-length channel with contacts near one FDW only. For ferroelectrics with extremely high permittivity, such as relaxors or PbZr$_x$Ti$_{1-x}$O$_3$ with composition x near the morphotropic phase boundary x=0.52, $G_+^{total}/G_-^{total}$ ratio can be essentially smaller then 1 for the case of a pronounced diffusive regime of current as well. This allows us to consider ultra-high permittivity ferroelectric substrates as excellent candidates for the fabrication of the novel rectifiers based on graphene p-n junction.

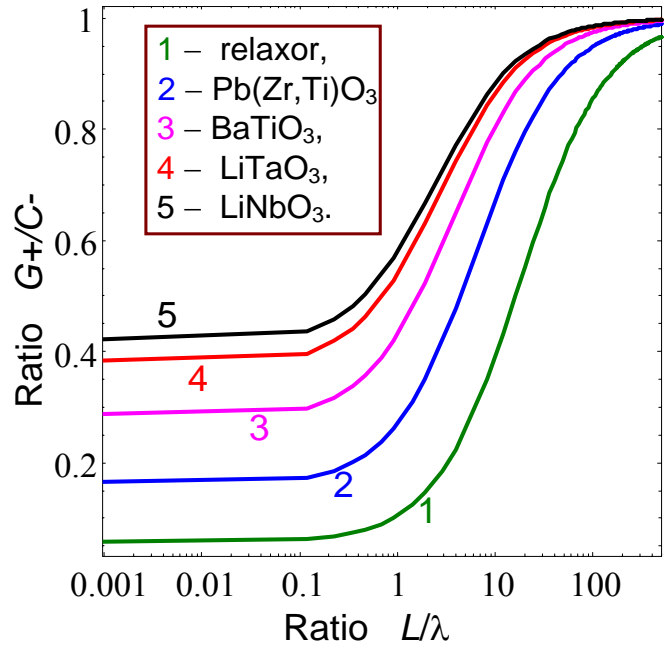

**Figure 2.3.** The ratio of the p-n junction total conductions for opposite polarities of applied voltages, $G_+^{total}/G_-^{total}$, as a function of the ratio of the graphene channel length to the mean path, $L/\lambda$, calculated at room temperature for different ferroelectric substrates, namely for several $\beta(\varepsilon_{33}^f)$, where the dielectric permittivity $\varepsilon_{33}^f \geq 5000$ for ferroelectric relaxor, $\varepsilon_{33}^f = 500$ for Pb(Zr,Ti)O$_3$, $\varepsilon_{33}^f = 120$ for BaTiO$_3$, $\varepsilon_{33}^f = 50$ for LiTaO$_3$ and $\varepsilon_{33}^f = 36$ for LiNbO$_3$ (curves 1-5). Reproduced from [M.V. Strikha, A. N. Morozovska. J. Appl. Phys. **120**, 214101 (2016)], with the permission of AIP Publishing.

## III. HYSTERETIC PHENOMENA IN GRAPHENE-ON-FERROELECTRIC
### A. Theoretical formalism

Generally the dependence of graphene channel conductivity on gate voltage is considered to be excellently symmetric one, $G(V_g) = G(-V_g)$ (Ref.[3]). This occurs, however, for the specially treated high quality graphene surfaces only. In many real imperfect structures with absorbed dipoles on a free graphene centers, localized states at graphene-substrate interface etc. this symmetry vanishes. Moreover, the dependence of $G(V_g)$ can get a hysteretic form (see e.g. Refs.[63-65] and Reviews [66, 67]). This hysteresis can have opposite directions of it's loop, and needs special treatment with allowance for different rival physical factors.

Let us consider a conducting graphene channel, placed on a dielectric or ferroelectric substrate (see Ref.[16] and **Fig. 3.1**). The concentration of 2D carriers in the channel is governed by several factors, which are a time-dependent gate voltage $V_g(t)$, polarization of the ferroelectric substrate dipoles, polarization of dipoles (e.g. absorbed water molecules) on graphene surface, and the trapped charge carriers localized at graphene-substrate interface. Further theoretical formalism are taken from the Ref.[16].

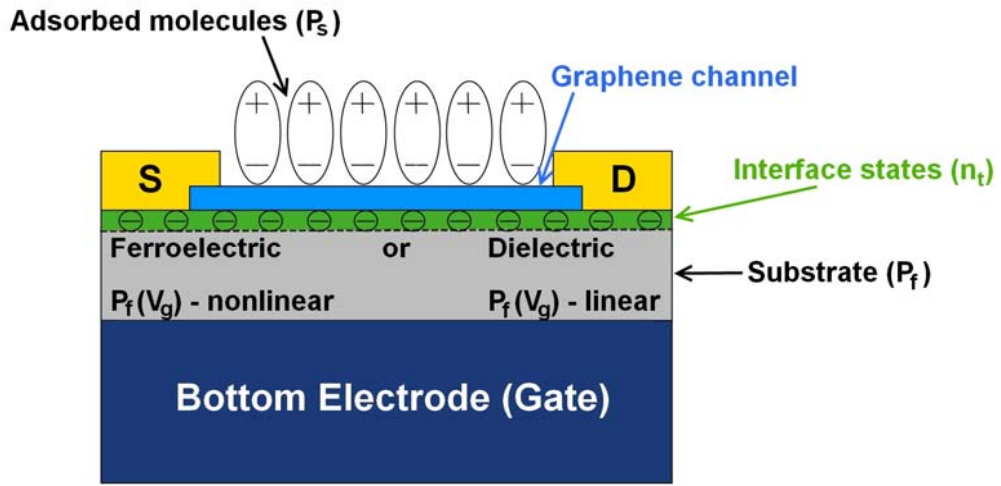

**Figure 3.1.** A graphene sheet (channel) placed on a dielectric or ferroelectric substrate. Dipoles (e.g. polar water molecules) can be absorbed by the graphene free surface. Carriers from the channel can be trapped by the centers localized at graphene-substrate interface. Reproduced from [A.I. Kurchak, A. N. Morozovska, M. V. Strikha. Journal of Applied Physics, **122**, 044504 (2017)], with the permission of AIP Publishing.

From all the abovementioned factors, the resistivity of graphene channel can be presented as [16]:

$$\rho[V_g, P_s, P_f, T] \approx \frac{1}{\sigma(V_g, P_s, P_f, T)} + \frac{1}{\sigma_{intr}(T)} + \frac{1}{\sigma_{min}}, \qquad (3.1)$$

where $P_s(t)$ is the polarization of dipoles absorbed on graphene surface, $P_f(t,T)$ is the temperature-dependent ferroelectric polarization. For the usual case of graphene under ambient conditions the scattering of ionized impurities by substrate dominate [38], and so the conductivity of 2D graphene channel is $\sigma(V_g, P_s, P_f, T) = e\mu n(V_g, P_s, P_f, T)$, where $n(V_g, P_s, P_f)$ is 2D carriers concentration per unit area, caused by the gate mixed doping, and by the dipoles absorbed by the surface, as well as by ferroelectric dipoles; $\mu$ is carriers mobility. The second term in Eq.(3.1) is the intrinsic graphene conductivity, $\sigma_{intr}(T) = e\mu n_{intr}(T)$, $n_{intr}(T) = \frac{2(k_B T)^2}{\pi(\hbar v_F)^2}$. The third term in Eq.(3.1) corresponds the minimal quantum conductivity, $\sigma_{min} \approx \frac{4e^2}{\hbar}$, that becomes significant at low $T$.

Tthe localized states are present at the graphene-substrate interface. For a voltage range $V_g(t) < V_{T1}$, where $V_{T1}$ corresponds to the situation when the occupation of the interface states with the electrons from graphene channel starts, $E_F(V_{T1}) = E_{T1}$, the 2D concentration of electrons in graphene channel is given by a capacitor formula [39, 40], $n(V_g, t) = \frac{\kappa V_g(t)}{4\pi e d}$, where $\kappa$ is the dielectric permittivity of substrate and $d$ is the substrate thickness. The gate voltage $V_{T1} = \frac{4\pi e d}{\kappa} \frac{E_{T1}^2}{\pi \hbar^2 v_F^2}$ leads to the start of the interface states occupation with electrons from the graphene channel. The voltage $V_{T2} = \frac{4\pi e d}{\kappa} \frac{E_{T2}^2}{\pi \hbar^2 v_F^2} + \frac{4\pi e d n_T}{\kappa}$ corresponds to the situation, when all localized interface states are already occupied by electrons from the graphene channel [ ].

In the voltage range $V_{T1} \leq V_g < V_{T2}$, for which the occupation of interface states occur, $n = \frac{E_{T1}^2}{\pi \hbar^2 v_F^2}$. In the voltage range $V_{T2} \leq V_g$ where all the interface states are already occupied by electrons, the free electrons' concentration in the channel is governed by the evident relation $n(V_g) = \frac{\kappa V_g(t)}{4\pi e d} - n_T$.

Let us take into consideration the polarization of dipoles in the ferroelectric substrate and at the graphene surface. If $E_F \leq E_{T1}$, similarly to [, ] we get:

$$n(V_g, P_s, P_f) = \frac{\kappa V_g(t)}{4\pi e d} + \frac{P_s(t) + P_f(t,T)}{e} \quad (3.2)$$

The temperature dependent spontaneous ferroelectric polarization can be described by expression [41]:

$$P_f(t,T) = P_f(T) \tanh\left(s_f \left(V_g(t) - V_c\right)\right), \quad (3.3)$$

where $V_c = E_c d$ is a coercive voltage equal to the product of coercive field and substrate thickness $d$, $s_f$ is the "sharpness" of ferroelectric switching. The spontaneous polarization of the dipoles absorbed by the graphene surface is:

$$P_s(t) = P_s \frac{1 - \tanh\left(s_s \left(V_g(t) - V_s\right)\right)}{2} \quad (3.4)$$

where $V_s$ is a critical voltage, that finally destroys the polarization; $s_s$ is a parameter reflecting the "sharpness" of dipoles switching. For the backward sweep polarizations $P_f$ and $P_s$ are

$$P_f(t,T) = P_f(T) \tanh\left(s_f \left(V_g(t) + V_c\right)\right) \text{ and } P_s(t) = P_s \left[1 - \exp\left(-\frac{t(V_s) - t}{\tau}\right)\right], \quad (3.5)$$

where $t(V_s)$ is the moment of time, corresponding to the complete suppression of $P_s$ by the critical gate voltage, $\tau$ is the dipoles' relaxation time that can be of several seconds [42]. Equations (3.3)-(3.5) are written for the case when absorbed dipoles recover immediately after the switching to the backward sweep. Above equations are valid for the case, when a lifetime of the carriers trapped by interface states is much greater than the switching time. The validity of this approximation for graphene on PZT substrate was demonstrated experimentally in Ref.[43].

The nonlinear response of ideal ferroelectric substrate can be described using Ginzburg-Landau-Khalatnikov relaxation equation [34, 44, 45]:

$$\Gamma \frac{dP(t)}{dt} = \alpha(T) P(t) + \beta P^3(t) + \gamma P^5(t) - E(\omega, t) \quad (3.6)$$

where $\alpha$, $\beta$ and $\gamma$ are Landau potential expansion coefficients.

**B. Gate control of the carrier concentration in the graphene channel**

*B1. Impact of the surface dipoles on the graphene carriers.* The dipoles can be absorbed by graphene surface (e.g. water molecules, studied experimentally [46, 47, 48]) and for this case $P_s \neq 0$, $P_f = 0$, $n_T = 0$. The dipoles shift the electro-neutrality point into the positive range of $V_g$ [16] and so the conductivity of graphene channel is determined by holes that at zero gate voltage. For this case the carrier concentration in graphene channel is determined both by gate doping and by the absorbed dipoles polarization, which in turn depends on the external electric field, caused by the gate voltage. The estimates [16] showed that the conductivity of the graphene channel would be governed by time-dependent polarization of the dipoles.

**Figure 3.2(a)** illustrates the dependence of carrier concentration on gate voltage in the graphene channel on SiO$_2$ substrate for different times of surface dipoles relaxation $\tau$. In the initial moment of time (at $V_g = 0$) graphene's channel conductivity is determined by holes, as it was noted before. However, the electric field, caused by the gate potential, destroys the dipoles polarization at some critical value of $V_g$. Solid lines correspond to the second forward sweep, i.e. polarization disappearance after the first cycle of its recover. The polarization will start to renew at the backward sweep. The value of polarization that recovers within a switching period is different for different relaxation times $\tau$, but its recovers almost completely within a switching period for a short relaxation time. The direction of such hysteresis loop would be opposite to the so-called direct one, created by ferroelectric dipoles ("anti-hysteresis").

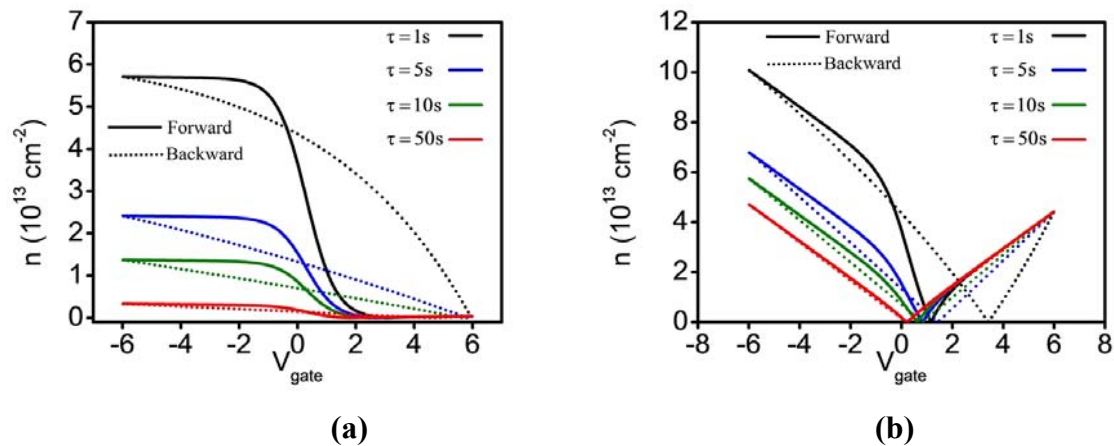

**Figure 3.2.** The dependence of carrier concentration in the graphene channel on SiO$_2$ substrate **(a)** and PZT substrate **(b)** vs. the gate voltage calculated for different times $\tau$=1, 5, 10 and 50 s (different colours) of surface dipoles relaxation. Forward and backward sweeping dependences are presented by solid and dotted curves respectively. The time dependence $V_g(t)$ is saw-like with a period $2\pi/\omega$ and the amplitude $A$. Reproduced from [A.I. Kurchak, A. N. Morozovska, M. V. Strikha. Journal of Applied Physics, **122**, 044504 (2017)], with the permission of AIP Publishing.

*B2. Impact of the ferroelectric substrate on the graphene carriers for low gate voltages.* Pb(Zr$_x$Ti$_{1-x}$)O$_3$ (PZT) is a material of choice for GFETs [, 49, 50], because its relative dielectric permittivity varies can be very high near the morphotropic phase boundary at $x = 0.52$. So that let us review the calculation results [16] of graphene channel conductance on PZT substrate. Taking into account that PZT response is not hysteretic for the electric fields smaller than the coercive field, $E_c$, (see Refs. [, ]), below we will operate within a relatively narrow gate voltage from – 8V to + 8V for which the linear response approximation $P(V_g)$ is valid. Therefore the further analysis should be similar to the one presented in B.1, but the ferroelectric substrate permittivity value is much higher than unity.

**Figure 3.2(b)** shows the carrier concentration in graphene calculated for different dipoles' relaxation times τ. The graphene channel on PZT substrate possess the hole conductivity at $V_g = 0$ similarly to the case of SiO$_2$ substrate. For positive gate voltages the holes concentration decreases and finally reaches the electro-neutrality point, where the graphene valence band is completely occupied by electrons, and the conduction band is empty. The graphene channel conductivity switches to an electron one with a further increase of the gate voltage.

*B3. Impact of the ferroelectric substrate on the graphene carriers for high gate voltages*

**Figure 3.3** presents the carrier concentration in graphene channel on ideal ferroelectric substrate, $n(V_g)$, as a function of the gate voltage at room temperature. **Fig.2.3** shows a pronounced hysteresis in $n(V_g)$ dependence corresponding to the ferroelectric polarization reversal induced by the change of the gate voltage polarity from the maximal negative value $-V_g^{max}$ to the positive one $+V_g^{max}$. In contrast to the previous case (shown in **Fig.2.2**) the hysteresis in **Fig.2.3** is caused by the two different stable ferroelectric polarization states and can be used for nonvolatile memory cells ' . However, corresponding switching times are comparative with the times of ferroelectric domains reversal imposing strict limitations on the operating frequency ($10^4$ - $10^6$)Hz for thick ferroelectric slabs. However, these times can be several orders of magnitude smaller (about $10^{-9}$s) for thin ferroelectric films of thickness less than 100 nm .

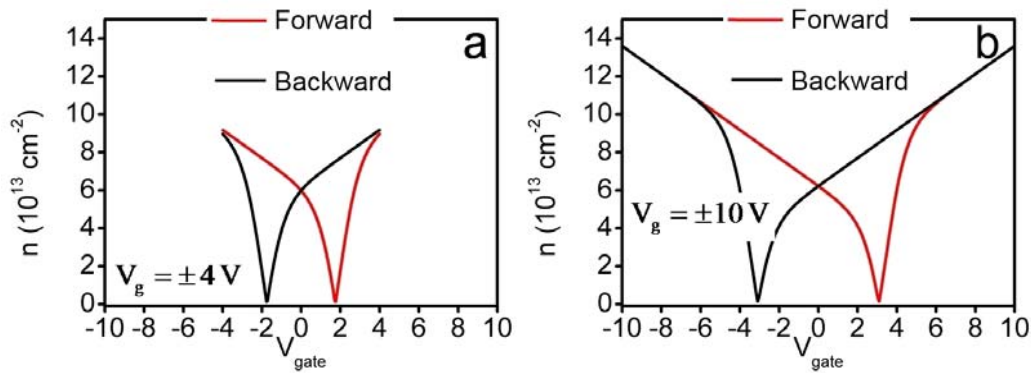

**Figure 3.3.** The dependence of carrier concentration in the graphene channel on ideal ferroelectric substrate vs. gate voltage calculated for different amplitudes of the gate voltage switching $V_g^{max} = \pm 4V$

(a) and $V_g^{max} = \pm 10V$ (b). Reproduced from [A.I. Kurchak, A. N. Morozovska, M. V. Strikha. Journal of Applied Physics, **122**, 044504 (2017)], with the permission of AIP Publishing).

*B4. Comparison with experiment for the realistic case.* To compare with realistic experiments let us consider two rival mechanisms, which control the carriers concentration in graphene, the first one originating from ferroelectric dipoles and the second one steaming from absorbed surface dipoles. **Figure 2.4** presents experimental [51] and theoretical results of electro-neutrality point positions as the function of gate voltage sweeping rate for GFET with grapene sheet on 140 nm $PbZr_{0.2}Ti_{0.8}O_3$ /60 nm $SrRuO_3$/ $SrTiO_3$ (001) heterostructure. From **Figure 2.4** the calculated values are in a reasonable agreement with experimental data . The discrepancy between the theoretical and experimental dependences can be caused by the interface states between graphene and ferroelectric, as well as by chemical doping of graphene during its fabrication.

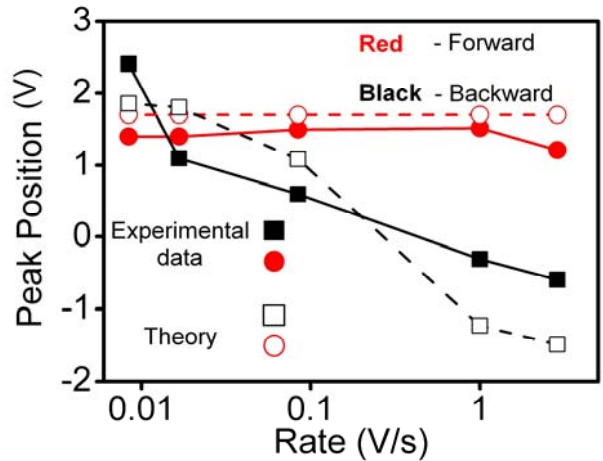

**Figure 3.4**. Experimental [ ] and theoretical values of electro-neutrality point positions as the function of gate voltage sweeping rate for GFET on PZT ferroelectric substrate Reproduced from [A.I. Kurchak, A. N. Morozovska, M. V. Strikha. Journal of Applied Physics, **122**, 044504 (2017)], with the permission of AIP Publishing.

We analyzed the origin of hysteretic form of the graphene conductivity dependence on a gate voltage for the graphene placed on different substrates. It was demonstrated, that the increase of the gate voltage sweeping rate $dV_g/dt$ leads to the disappearance of hysteresis for GFET on dielectric or ferroelectric substrates for the gate voltages less than the coercive one, $V_g<V_c$. The increase of gate voltage rate causes the transition from anti-hysteresis to ferroelectric hysteresis. These results are in qualitative and quantitative agreement with the experimental data .

Notice that results of Ref.[16] have been obtained using several approximations and simplifications, which do not account for concrete physical nature of dipoles bonding to graphene surface and for possible relaxation time spectrum, as well as the carriers trapping by surface states are dependent on the frequency of gate voltage switching in reality. The understanding of the effects is possible on the base of the clear vision of the physical mechanism of trapping and require further studies.

## IV. DYNAMICS OF P-N JUNCTIONS IN GRAPHENE CHANNEL INDUCED BY THE MOTION OF FERROELECTRIC DOMAIN WALLS

**A. Theoretical formalism**

In Section II we considered the case of the fixed FDW. However, generally FDWs can move under external effects, which needs special examination. The review of theoretical formalism presented in this section is based on Ref.[17]. The typical geometry of the GFET with 2D-graphene layer (channel)-on-ferroelectric film with 180-degree **FDW** is shown in **Fig. 4.1.** Top gate is deposited on oxide layer and the graphene sheet is separated from the ferroelectric film by ultra-thin paraelectric dead layer, originated due several reasons, such as incomplete polarization screening at the surface and the imperfect deposition process of graphene on the ferroelectric. There are different types of such layers treated as ultrathin and located under the surface of ferroelectric substrate, where the spontaneous polarization is absent (or negligibly small) due to the surface contamination, zero extrapolation length and/or strong depolarization field [52, 53]. Periodic voltage applied to the top gate can induce the motion of FDWs in the ferroelectric substrate.

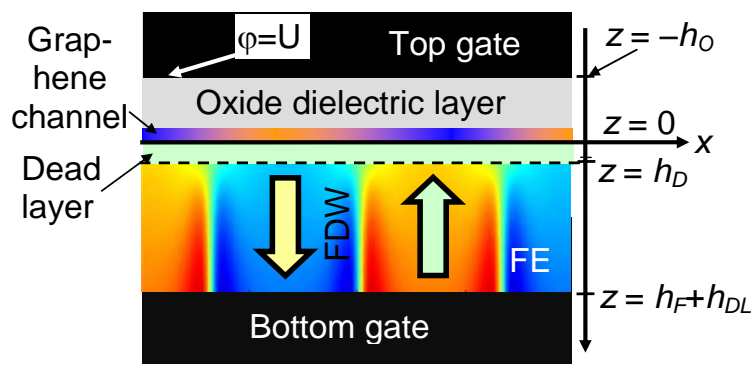

**Figure 4.1.** Schematics of the 180°-ferroelectric domain walls (FDWs) in the heterostructure "top gate – oxide dielectric layer – graphene channel – paraelectric dead layer – ferroelectric film – bottom gate". The ferroelectric substrate is in a perfect electric contact with the bottom gate electrode. Adapted from [A. I. Kurchak, et al. Phys. Rev. Applied **8**, 024027 (2017)], with the permission of APS Publishing for the authors artwork.

A single-layer graphene is regarded infinitely thin sheet with two-dimensional (2D) electron density of electrons and holes states, $g_n(\varepsilon) = g_p(\varepsilon) = 2\varepsilon/(\pi\hbar^2 v_F^2)$ (see e.g. [3, 40]). Hence the 2D concentration of electrons in the conduction band and holes in the valence band of graphene are $n_{2D}(\varphi) = \int_0^\infty d\varepsilon g_n(\varepsilon) f(\varepsilon - E_F - e\varphi)$ and $p_{2D}(\varphi) = \int_0^\infty d\varepsilon g_p(\varepsilon) f(\varepsilon + E_F + e\varphi)$, respectively, $E_F$ is a Fermi energy level. The graphene charge density is equal to $\sigma_G(\psi) = e(p_{2D}(\psi) - n_{2D}(\psi))$, where $\psi = \dfrac{e\varphi + E_F}{k_B T}$.

Equations of state $\mathbf{D} = \varepsilon_0 \varepsilon_O \mathbf{E}$ and $\mathbf{D} = \varepsilon_0 \varepsilon_{DL} \mathbf{E}$ relate the electrical displacement $\mathbf{D}$ and electric field $\mathbf{E}$ in the oxide dielectric and ultrathin dead layers of thicknesses $h_O$ and $h_{DL}$, respectively, $\varepsilon_0$ is a universal dielectric constant. The relative dielectric permittivity of the dead layer $\varepsilon_{DL}$ is rather high $\sim 10^2$ [54]. The potential $\varphi_{DL}$ satisfies Laplace's equation inside the dead layer. Note that the problem of dielectric permittivity of 2D-graphene layer is still under debate (see e.g. [55]).

A ferroelectric film has thickness $l$ and ferroelectric polarization $P_3^f$ directed along its polar axis z, with 180-degree domain wall – surface junctions [see **Fig. 4.1**]. Polarization z-component is $P_3(\mathbf{r}, E_3) = P_3^f(\mathbf{r}, E_3) + \varepsilon_0(\varepsilon_{33}^b - 1)E_3$, where a so-called relative "background" permittivity $\varepsilon_{ij}^b$ is introduced [52]. The values of $\varepsilon_{ij}^b$ are not related with a soft ferroelectric mode and limited by the linear dielectric response of the lattice, so for the most of ferroelectrics perovskites they are within the range (4 – 7) (see Ref.[56] for its determination and refs therein). The spatial distribution of the ferroelectric polarization $P_3(x, y, z)$ is determined from the time-dependent LGD type Euler-Lagrange equation,

$$\Gamma \frac{\partial P_3}{\partial t} + aP_3 + bP_3^3 + cP_3^5 - g\Delta P_3 = E_3. \quad (4.1)$$

$\Gamma$ is a Landau-Khalatnikov relaxation coefficient [57], and $g$ is a gradient coefficient, $\Delta$ stands for a 3D-Laplace operator. Constants $a = \alpha_T(T - T_C)$, $b$ and $c$ are the coefficients of LGD potential expansion on the polarization powers (also called as linear and nonlinear dielectric stiffness coefficients). Corresponding boundary conditions are of the third kind [58], with the extrapolation lengths $\Lambda_\pm$ [59].

For the problem geometry shown in the **Fig. 4.1** the system of electrostatic equations acquires the form listed in Ref.[17]. Boundary conditions to the system are fixed potential at the top ($z = -h_O$) and bottom ($z = h_{DL} + h_F \approx h_F$) gate electrodes; the continuity of the electric potential at the graphene layer ($z = 0$) and the equivalence of difference of the electric displacement normal components, $D_3^O = \varepsilon_0 \varepsilon_O E_3$ and $D_3^{DL} = \varepsilon_0 \varepsilon_{DL} E_3$, to the surface charges in graphene $\sigma_G(x, y)$; and the continuity of the displacement normal components, $D_3^f = \varepsilon_0 \varepsilon_{33}^b E_3 + P_3^f$ and $D_3^{DL} = \varepsilon_0 \varepsilon_{DL} E_3$, at dead layer/ferroelectric interface. The gate voltage is periodic with a period $T_g$, $U(t) = U_{max} \sin(2\pi t / T_g)$.

To generate moving domains ferroelectric film thickness should be above the critical thickness $l_{cr}$ of the size-induced phase transition into a paraelectric phase, at that $l_{cr}$ depends on the dielectric and dead layer thicknesses [60, 61, 62]. The domains appear above the critical thickness, since they minimize the depolarization field energy in the gap and dielectric layer [63].

Since the lateral dimension of a ferroelectric film $L_{FE}$ is typically much higher than the graphene channel length $L$, **odd**, **even** or **not fractional** number of domain walls can pass along the channel during the period of the gate voltage $T_g$ depending on the interrelation between the graphene channel length $L$ and the period $T_{FE}$ of the domain structure in a ferroelectric film (see Ref. [17]). The realistic situations can be modelled by a **periodic, mixed or antiperiodic** boundary conditions (**BCs**) on polarization component $P_3$, its derivative $\frac{\partial P_3}{\partial x}$, electric potential $\varphi_f$ and its derivative $\frac{\partial \varphi_f}{\partial x}$ at the lateral boundaries $x = \pm L/2$ [17].

Using the results [15, 17], for the *even* number (2k) of walls between source and drain electrodes of graphene channel its conductions $G_+^{total}$ and $G_-^{total}$ are equal for both polarities of the gate voltage $\frac{G_+^{total}}{G_-^{total}} = 1$, because for each polarity there are $k$ p-n junctions with conduction, given by eq.(8) of [15], and $k$ p-n junctions with conduction, given by eq.(10) of [15]. For the *odd* number of the walls $2k+1$, Eq.(14) of Ref.[15] can be modified as:

$$\frac{G_+^{total}}{G_-^{total}} = \frac{\beta(L+\lambda) + \lambda k(1+\beta)}{\beta(L+\lambda) + \lambda k(1+\beta) + \lambda}, \qquad (4.2)$$

where the factor $\beta = \sqrt{\frac{\pi \alpha c}{4 \varepsilon_{33}^f v_F}}$ is proportional to the squire root of the fine structure constant α, light velocity in vacuum *c*, and inversely proportional to the relative ferroelectric permittivity $\varepsilon_{33}^f$ and Fermi velocity of electrons in graphene $v_F$. The electron mean free path λ is typically essentially smaller that the channel length *L*.

For a pronounced diffusion regime of current, $\beta L \gg \lambda$, as well as for the great number of the walls, $k \gg k_{cr}$, where $k_{cr} = 1 + \frac{\beta L}{(1+\beta)\lambda}$, a conduction of graphene channel is given by an expression [3, 40], $G = \frac{\lambda(n_{2D})}{L}\frac{2e^2}{\hbar\pi^{3/2}}W\sqrt{n_{2D}}$, being proportional to the product of the graphene channel width $W$ on the squire root of 2D charge concentration $n_{2D}$. The mean free path $\lambda(n_{2D}) \sim \sqrt{n_{2D}}$ for the scattering at ionized centers in the substrate and in the temperature range far from Curie temperature [3].

**B. Analyses of ferroelectric domains and p-n junctions correlated dynamics**

Below we present results of numerical modeling of the considered problem. Parameters used in the calculations are listed in **Table I** in Ref. [17]. The polarization component in the ferroelectric film $P_3$, variation of 2D-concentration of free carriers in the graphene channel $\Delta n_G = (p_{2D} - n_{2D})$, and the effective conductance ratio $\Delta\eta(U_{max}) = \frac{\Delta n_G(+U_{max})}{\Delta n_G(-U_{max})}$ were calculated in dependence on the periodic gate voltage $U(t) = U_{max}\sin(2\pi t/T_g)$.

The hysteresis loops of the average polarization $P_3(U)$ and concentration variation $\Delta n_G(U)$ are shown in **Figs.4.2.** At relatively low voltages ($\leq 2$ V) polarization and concentration loops of quasi-elliptic shape [see black curves in **Fig. 4.2(a)-(d)**]. This happens because the film state is poly-domain, and the domain walls are moving by the electric field with the gate voltage changing from $-U_{max}$ to $+U_{max}$. With the increase of the gate voltage amplitude $U_{max}$ to (5 – 10) V the domain walls start to collide and the domains with opposite polarization orientation almost "annihilate" for definite periodic moments of time $t$, and then the polar state of the film with some degree of unipolarity partially restores [see red and magenta curves in **Figs.4.2(a)-(d)**]. The loops asymmetry and rectification ratio is defined by the symmetry of the BCs for polarization and electric potential at lateral surfaces $x = \pm L/2$.

Completely symmetric loops of polarization $P_3(U)$ and concentration $\Delta n_G(U)$ variation correspond to the periodic BCs [**Figs.4.2(a)** and **4.2(c)**]. The rectification effect is absent in the case of periodic BCs, since the effective ratio $\Delta\eta(U_{max}) \equiv -1$ for all $U_{max}$ [see **Figs.4.2(e)**], because the even number of domain walls is moving in the ferroelectric substrate at any time. Characteristic distributions of the polarization component $P_3$ and free charge concentration along the graphene channel are shown in **Figs. 4.3(a)** and **4.3(c)**, respectively. Two p-n junctions are induced by the two moving domain walls at certain times over one period.

The antiperiodic BCs lead to asymmetric loops of polarization $P_3(U)$ and concentration $\Delta n_G(U)$ [**Figs.4.2(b)** and **4.2(d)**]. The vertical asymmetry and horizontal shift of $P_3(U)$ loop are much stronger than the asymmetry of $\Delta n_G(U)$ loop because the polarization acts on the charge indirectly via the depolarization field. The asymmetry of $P_3(U)$ and $\Delta n_G(U)$ becomes weaker with maximal voltage increase [compare different loops in **Figs.4.2(c)** and **4.2(f)**]. The rectification effect is evident for low and moderate voltages, since the effective ratio $\Delta\eta(U_{max}) \ll 1$ for $1 < U_{max} < 4$, and the ratio saturates, $\Delta\eta(U_{max}) \to -1$, for higher voltages [see **Figs.4.2(f)**], because the odd number of domain walls is moving in the ferroelectric film most of the time. Characteristic distributions of the polarization and graphene charge are shown in **Figs. 4.3(b)** and **4.3(d)**, respectively. Three p-n junctions are induced by the three moving domain walls over one period.

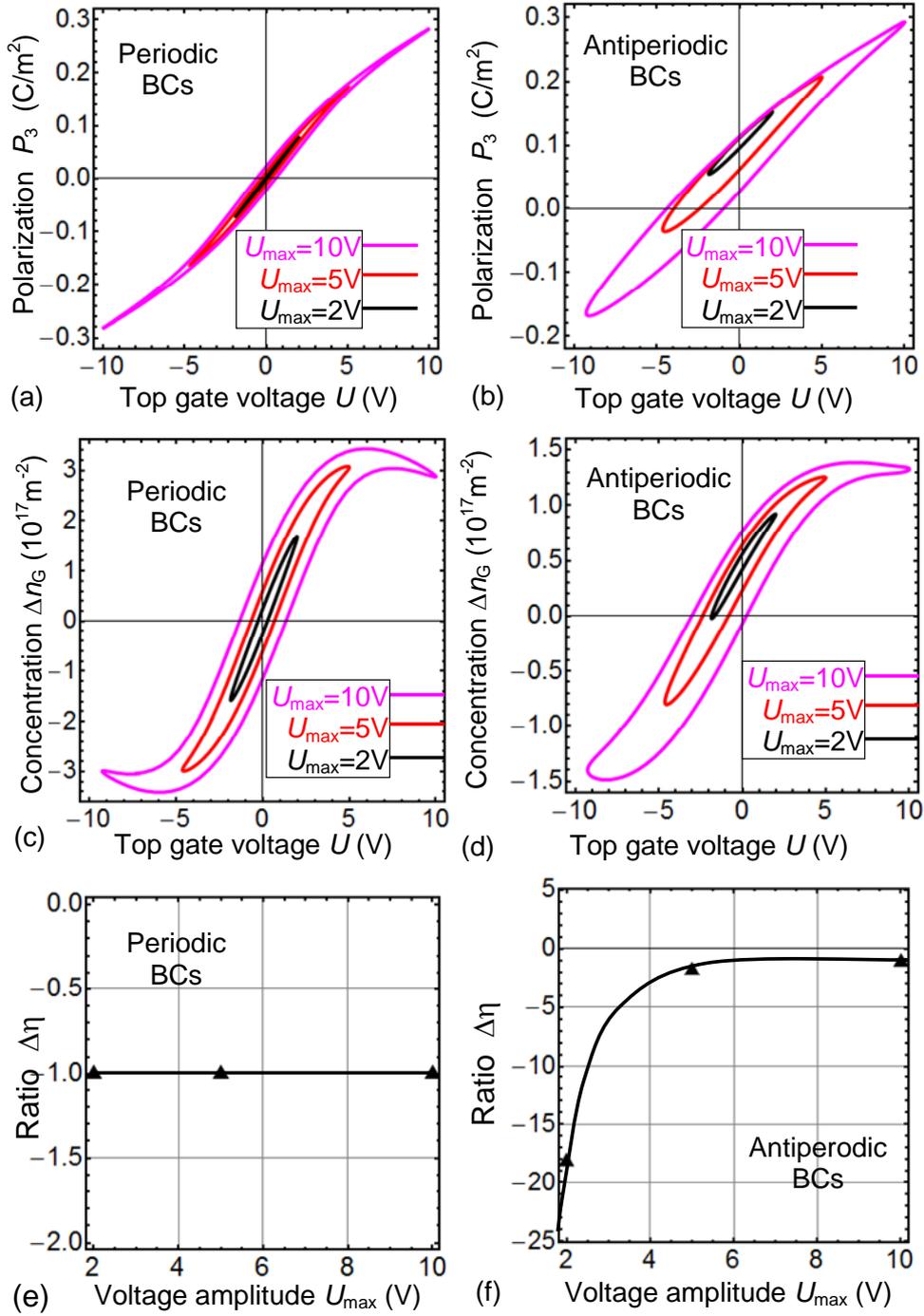

**Figure 4.2.** Hysteresis loops of ferroelectric polarization $P_3(U)$ **(a, b)**, carriers concentration variation in a graphene channel $\Delta n_G(U)$ **(c, d)** and the conductance ratio $\Delta\eta(U_{max})$ **(e, f)** calculated for periodic **(a, c, e)** and antiperiodic **(b, d, f)** boundary conditions (BCs). Black, red and magenta loops correspond to the different amplitudes of gate voltage $U_{max}$ = (2, 5, 10) V and gate voltage period $T_g=10^3$ s. Adapted from [A. I. Kurchak, et al. Phys. Rev. Applied **8**, 024027 (2017)] with the permission of APS Publishing for the authors artwork.

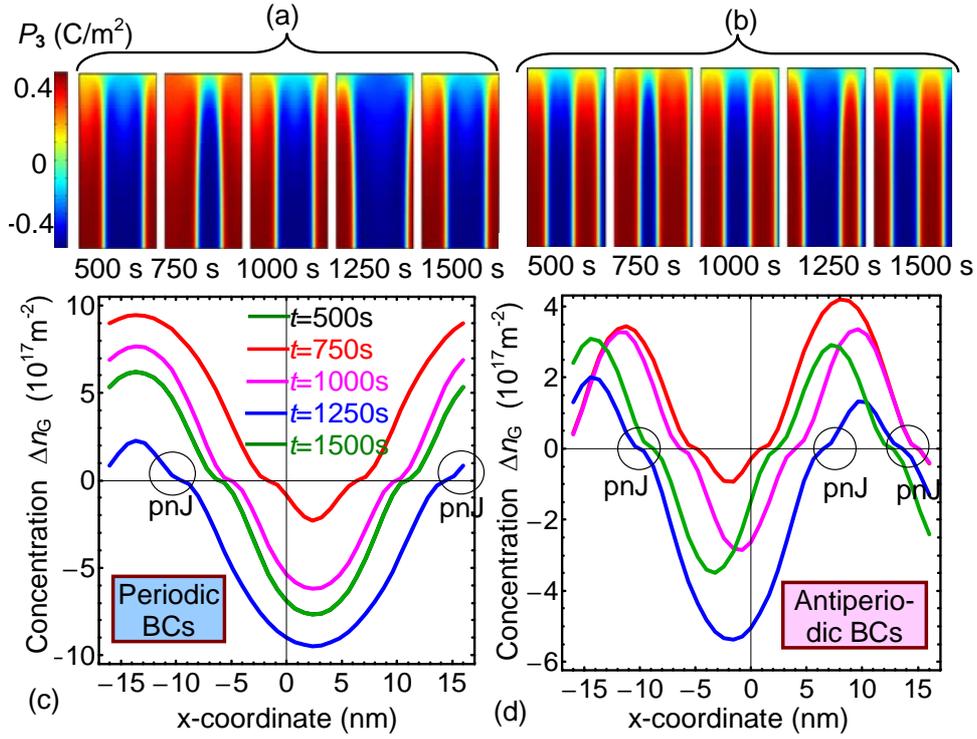

**Figure 4.3.** Spatial distribution of polarization component $P_3$ in a ferroelectric film calculated at certain times over a period, 500, 750, 1000, 1250 and 1500 s (specified in the plots) for the periodic **(a)** and antiperiodic **(b)** BCs. 2D-concentration of the free charge $\Delta n_G(x,t)$ calculated along the graphene channel at certain times over a period, 500, 750, 1000, 1250 and 1500 s (specified in the plots) for the periodic **(c)** and antiperiodic **(d)** BCs. Gate voltage amplitude $U_{max} = 5$ V and period $T_g = 10^3$ s. Other parameters are the same as in **Fig.3.2.** The transient process that almost vanish enough rapidly is not shown (so that we started to show the plots from times $t \geq 500$ s). Adapted from [A.I. Kurchak, et al. Phys. Rev. Applied **8**, 024027 (2017)] with the permission of APS Publishing for the authors artwork.

### C. Extrinsic size effect

Note that the conductivity of graphene channel (that is proportional to the total carrier concentration variation $\Delta n_G(L,t)$) depends on its length $L$ [see the contour map of the average concentration variation $\Delta n_G(L,t)$ in coordinates "channel length $L$ – time $t$" in **Figs. 4.4**]. The phenomenon we called "extrinsic size effect" [15] consists in a quasi-periodic modulation of the conductivity amplitude with the channel length for periodic BCs, at that the modulation extremes are the most pronounced for $L$ around 27.5 nm (the first maxima and minima depending on the time moment) and 50 nm (the second maxima and minima) at the parameters listed in Ref. [15]. The modulation becomes less pronounced with $L$ increase [compare the contrast for the modulation maxima and minima at $L$=27.5 nm and 50 nm]. Also the distance $\Delta L$ between the maxima is voltage-independent and slightly increases with $L$ increase. We expect that the extrinsic size effect will disappear for long channels, which length is much longer than the intrinsic period of domain structure in a ferroelectric substrate

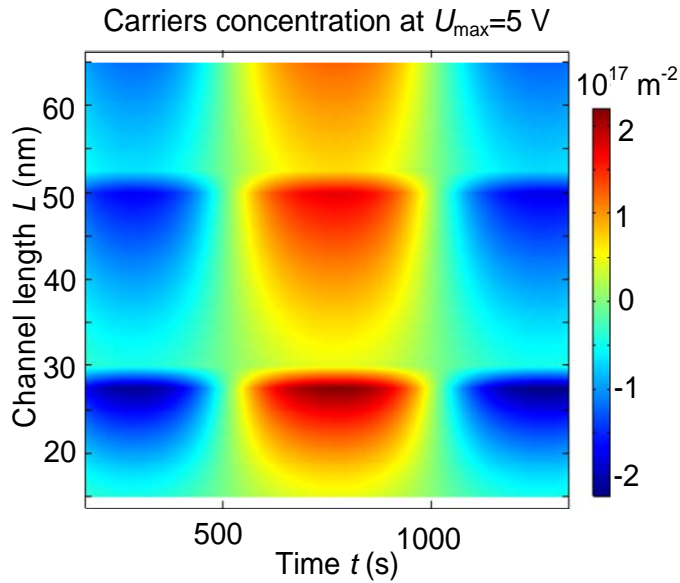

**Figure 4.4.** Dependence of the graphene channel conductivity on its length and time calculated for the periodic BCs at $U_{max}$ = 5 V. Adapted from [A. I. Kurchak, et al. Phys. Rev. Applied **8**, 024027 (2017)] with the permission of APS Publishing for the authors artwork.

## V. GRAPHENE SEPARATION AND STRETCHING INDUCED BY PIEZOELECTRIC EFFECT OF FERROELECTRIC DOMAINS

P-n junctions in graphene on ferroelectric substrates have been actively studied, but the impact of the piezoelectric effect in ferroelectric substrate with ferroelectric domain walls (FDWs) on graphene characteristics was considered only recently. Namely, as it has been shown in Ref.[18] the elastic strain can significantly affect the graphene conductance via the ***stretching*** of its surface and ***separation*** of graphene areas at the steps between elongated and contracted domains. The review of theoretical formalism presented in this section is based on Ref.[18].

The idea is that ferroelectric domain stripes with opposite spontaneous polarizations elongate or contract depending on the polarity of voltage applied to the substrate due to the piezo-effect. If the voltage is applied to a gate of the GFeFET with FDW, one domain elongates and another one contracts depending on the voltage polarity [see **Fig. 5.1**]. The surface displacement can be significant for ferroelectrics with high piezoelectric coefficients, such as PbZr$_{0.5}$Ti$_{0.5}$O$_3$ (**PZT**) those piezoelectric coefficients can reach (0.3 – 1)nm/V depending on the film thickness and temperature [64]. Corresponding piezoelectric surface displacement step $h$ is about (0.5 – 1) nm for the gate voltage ~(1 – 3)V. The thickness $d \leq 0.5$ nm of the physical gap between the graphene and ferroelectric is determined by Van-der-Waals interaction. The density $J$ of binding energy for graphene on SiO$_2$ substrate is about 0.5 J/m$^2$ [65]. Because graphene adhesion to SiO$_2$ should be the strongest one in comparison with other surfaces, it is natural to expect that $J$ value for graphene on PZT surface is smaller. The Young's modulus Y of graphene is giant, 1 TPa, [66, 67]. Under these conditions the partially separated graphene region occurs at the step when the normal component $F_n$ of the elastic tension force $F$ applied to a carbon atom exceeds the force $F_b$ binding the atom with the surface [see the scheme of the forces in **Fig. 5.1(b)**]. The separated section of length $l + \Delta l$ is "*suspended*" between the bounded sections.

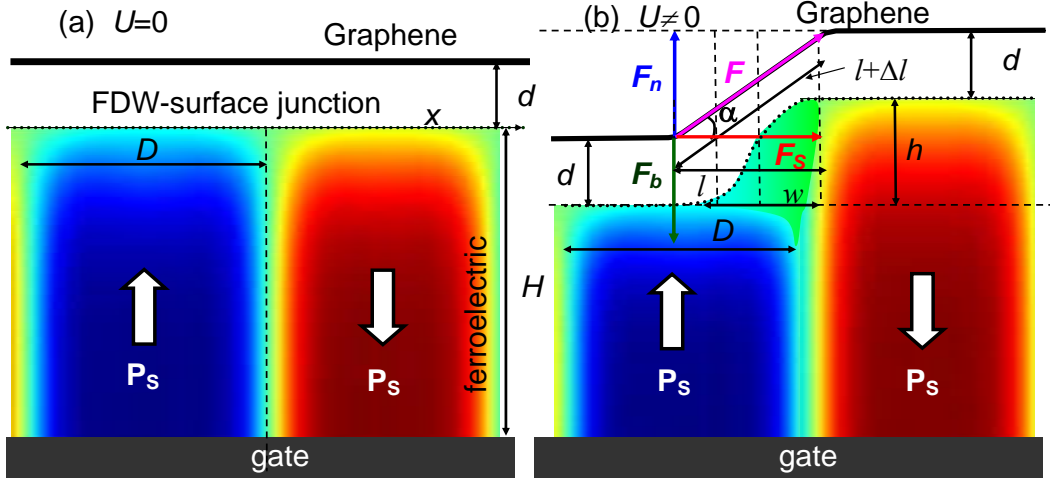

**Figure 5.1.** Partial separation of graphene channel sections induced by a piezoelectric effect at the ferroelectric domain wall – surface junction. The separation is absent at zero applied voltage $U=0$ **(a)** and appears at nonzero $U \neq 0$ **(b)**. $F$ is the elastic tension force, $F_n$ is its normal component, $F_S$ is its lateral component and $F_b$ is the binding force of the carbon atom to the surface. Adapted from [A. N. Morozovska, A. I. Kurchak, .M. V. Strikha. Phys. Rev. Applied **8**, 054004 (2017)] with the permission of APS Publishing for the authors artwork.

**A. Piezoelectric displacement of the ferroelectric substrate surface**

Analytical expression for the vertical displacement $u_3(x)$ in the vicinity of FDW-surface junction is derived in Ref.[18] within the framework of decoupling approximation [68, 69, 70, 71, 72]. The displacement $u_3(x)$ acquires the form [18]:

$$u_3(x) = -U[W_{33}(x)d_{33} + W_{31}(x)d_{31}]. \tag{5.1}$$

The $U$ is the voltage between the top and bottom electrodes, i.e. the gate voltage; $d_{33}$ and $d_{31}$ are piezoelectric coefficients. The concrete forms of $W_{33}(x)$ and $W_{31}(x)$ are given in Ref.[18].

**Figure 5.2** shows the typical profiles of the PZT surface displacement calculated for the gate voltage $U=1$ V and room temperature. Values of the PZT film thickness $H$ in the range $H = (20 - 500)$ nm and separation $d = 0.5$ nm. One can see from the figure that the step originated at FDW-surface junction is the widest for the smallest ratio $d/H$ (curve 1), and becomes essentially thinner with the ratio increase (curves 2-4). As one can see, the maximal height $h$ of the step changes in a non-monotonic way with $d/H$ ratio increase, but the displacement difference far from the domain wall, $h_\infty = |u_3(x \to \infty) - u_3(x \to -\infty)|$, is the same for all curves and is given by expression [18]:

$$h_\infty = 2|U|[d_{33} + (1 + 2\nu)d_{31}]. \qquad (5.2)$$

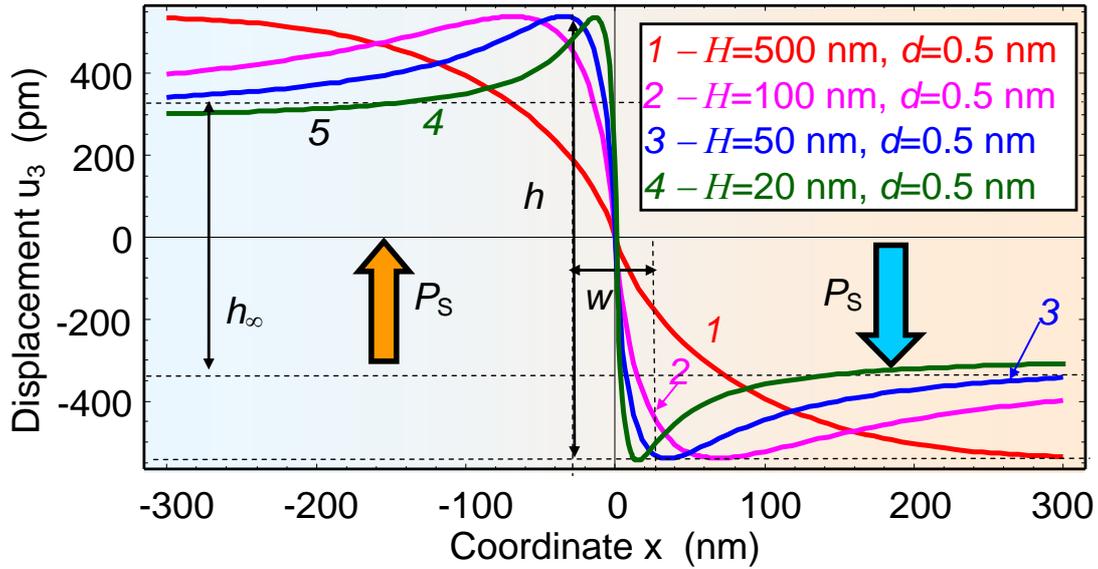

**Figure 5.2.** Profiles of the ferroelectric surface displacement $u_3$ calculated at the gate voltage U=1 V, thermodynamic piezoelectric coefficients $d_{33} \approx 10^3$ pm/V, $d_{31} \approx -450$ pm/V and Poisson ratio $\nu$=0.3 corresponding to PZT at room temperature. The values of ferroelectric film thickness $H$ and distance $d$ for the curves 1 - 4 are indicated in the legend. Note that vertical picometer scale is much smaller than the horizontal nanometer scale. Adapted from [A. N. Morozovska, A. I. Kurchak, .M. V. Strikha. Phys. Rev. Applied **8**, 054004 (2017)] with the permission of APS Publishing for the authors artwork.

The complete separation (i.e. exfoliation) of graphene caused by a piezoelectric effect is hardly possible in thick ferroelectric films with smooth profiles of the surface displacement corresponding to the curve 1 in **Fig.5.2**. The complete exfoliation becomes impossible if the domain wall is a single one in a film; and so only the stretching of graphene sheet is induced by the piezoelectric effect in the case. The partial separation or complete exfoliation of graphene can be favorable when it contacts with relatively sharp ferroelectric surface profiles across the FDWs corresponding to the curves 2-4 in **Fig.5.2**.

## B. Graphene separation, stretching and exfoliation: impact on the conductance

Taking into that the normal component $F_n$ of the tangential force $F$ and binding force $F_b$ are equal [see **Fig.5.1(b)**], we derived analytical expression for the minimal length $l$ of separated graphene region [18],

$$l = h \cdot \sqrt[3]{\frac{Yd}{2J}} > |U|(d_{33} + (1+2\nu)d_{31})\sqrt[3]{\frac{4Yd}{J}}. \quad (5.3)$$

The inequality in Eq.(5.3) originates from the inequality $h > h_\infty$ [see **Fig.5.2**]. Estimates made from Eq.(5.3) give that the stretched section can reach tens of nm for gate voltages $|U| \geq 3$ V, PZT parameters at room temperature, binding energies $J < 0.25$ J/m$^2$ and separation $d$=0.5 nm.

The conductance of graphene channel in diffusion regime can change significantly, because electrons in the separated stretched section scatter on acoustic phonons [40]. The voltage dependence of the conductance $G(U)$ of graphene channel of length $L$, when its part of length $l(U)$ is separated and another part of length $L - l(U)$ is bounded, obeys the Matiessen rule [40]:

$$G(U) = W\left[\frac{L-l(U)}{\sigma_B} + \frac{l(U)}{\sigma_S}\right]^{-1} \quad (5.4)$$

The separated length $l(U) = \chi|U|$, where the coefficient $\chi = (d_{33} + (1+2\nu)d_{31})\sqrt[3]{4Yd/J}$ in accordance with Ref. [18].

The conductivity of the bounded section has the form [3]:

$$\sigma_B = \frac{2e^2}{\pi^{3/2}\hbar}\lambda_B\sqrt{n_{2D}}. \quad (5.5)$$

Here $e$ is elementary charge, $\hbar$ is Plank constant, $v_F = 10^6$ m/s is characteristic electron velocity in graphene, $\lambda_B$ is mean free path in the bounded part of graphene channel. The concentration of 2D electrons $n_{2D}$ can be regarded constant voltage-independent value far from the FDWs, namely $n_{2D} \approx |P_S/e|$ [17, 18, 34]. Using that for the most common case of electron scattering in graphene channel at ionized impurities in a substrate $\lambda_B[n_{2D}] = \xi\sqrt{n_{2D}}$, where the proportionality coefficient $\xi$ depends on the substrate material and graphene-ferroelectric interface chemistry, we obtain from Eq.(5.5) the dependence $\sigma_B[n_{2D}] = \frac{2e^2\xi}{\pi^{3/2}\hbar}n_{2D} \approx 8.75 \cdot 10^{-5}\xi n_{2D}$ (in Siemens). Taking into account that $P_S$ value can be 10 times smaller for thin films than its bulk value, the concentration vary in the range $n_{2D} \cong (0.3-3) \times 10^{18}$ m$^{-2}$ depending on the film thickness, but should be regarded voltage- and coordinate- independent constant far from the FDW. Thus elementary estimates give $\sigma_B \cong (0.15 - 15) \times 10^{-3} \Omega^{-1}$ for reasonable ranges of $\lambda_B = (10 - 100)$ nm and $P_S = (0.05 - 0.5)$ C/m$^2$.

On the contrary, the main channel for electron scattering in the separated stretched section of structurally perfect graphene is collisions with acoustic phonons. In this case $\lambda_S(E) \sim 1/E$ [3]. This leads to a well known paradox: conductivity $\sigma_S$ doesn't depend on 2D electrons concentration in the graphene channel. Hence for estimations a well known upper limit for $\sigma_S$ [3] can be used:

$$\sigma_S = \frac{4e^2 \hbar \rho_m v_F^2 v_S^2}{\pi D_A^2 k_B T} \qquad (5.6)$$

Here $\rho_m \approx 7.6 \cdot 10^{-7}$ kg/m$^2$ is 2D mass density of carriers in graphene, $v_S \approx 2.1 \cdot 10^4$ m/s is a sound velocity in graphene, Boltzmann constant $k_B = 1.38 \times 10^{-23}$ J/K; $D_A \approx 19$ eV is acoustic deformation potential that describes electron-phonon interaction. Expression (5.6) yields $\sigma_S \approx 3.4 \times 10^{-2} \Omega^{-1}$ at room temperature.

**Figures 5.3** presents the conductance $G$ calculated for different values of gate voltage $U$, binding energy $J$ and channel length $L$. The conductance increases with $U$ increase; at that the increase is monotonic and faster than linear and the most pronounced at small binding energies $J \leq 0.2$ J/m$^2$ [**Fig. 5.3(a)**] and small channel length $L \leq 100$ nm [**Fig. 5.3(b)**]. The conductance ratio $G(U)/G(0)$ doesn't exceed 1.25 for the case of graphene partial separation [presented by the red curves] at realistic values of parameters. However, it can be significantly greater, if the domain stripe period $D$ is much shorter than the channel length $L$. For the case length $L$ is divided in two almost equal parts between the separated and bonded sections (i.e. $l \approx L/2$). If p-n-junctions at FDW don't change the general conductance of the graphene channel significantly, the electron mean free passes $\lambda_B \ll D$, $\lambda_S \ll D$ and $\sigma_S \gg \sigma_B$, Eq.(5.4) yields $\frac{G(U)}{G(0)} \approx 2$. Note that the ratio $G(U)/G(0)$ can be significantly greater than 2, e.g. in the case of mostly suspended graphene ($l \approx L$) with $\sigma_S \gg \sigma_B$. However, the possibility of such a limiting case needs special examination.

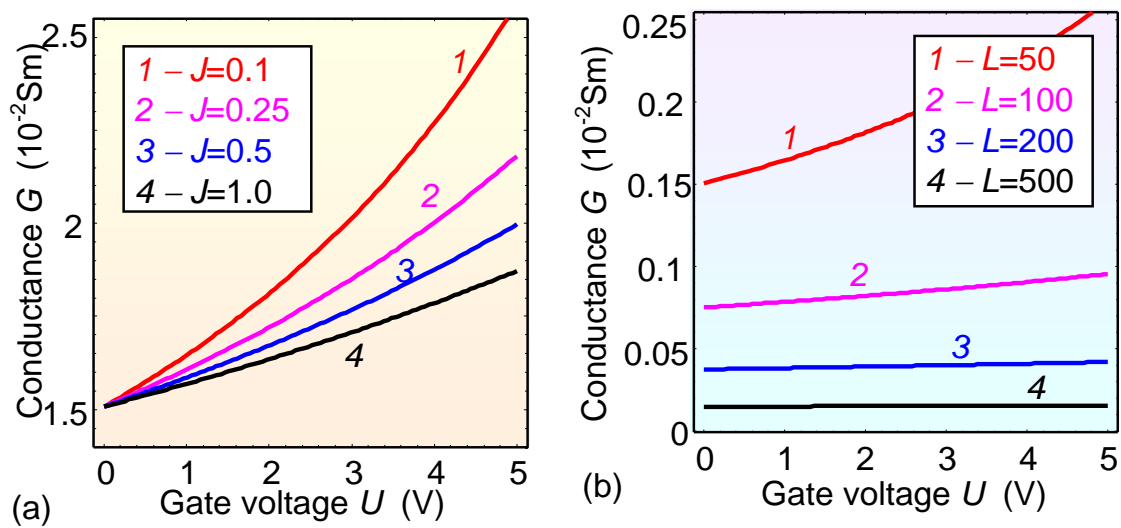

**Figure 5.3.** Dependences of the conductance $G(U)$ on the gate voltage $U$ calculated for several values (curves 1 – 4) of binding energy $J$=0.1, 0.25, 0.5, 1.0 J/m$^2$ **(a)** and channel length $L$ = 50, 100, 200, 500 nm **(b)**. Adapted from [A. N. Morozovska, A. I. Kurchak, .M. V. Strikha. Phys. Rev. Applied **8**, 054004 (2017)] with the permission of APS Publishing for the authors artwork.

**Figure 5.4** schematically illustrates the conductance $G(U)$ calculated from Eqs.(5.4)-(5.6). The predicted effect of conductance modulation can be very useful for improvement and miniaturization of various electronic devices (such as advanced logic elements, memory cells, high efficient hybrid electrical modulators and transducers of voltage-to-current type, and piezo-resistive elements). Also we propose the alternative method of suspended graphene areas fabrication based on the piezo-effect in a ferroelectric substrate. The method does not require any additional technological procedures like chemical etching or mechanical treating of the substrate surface.

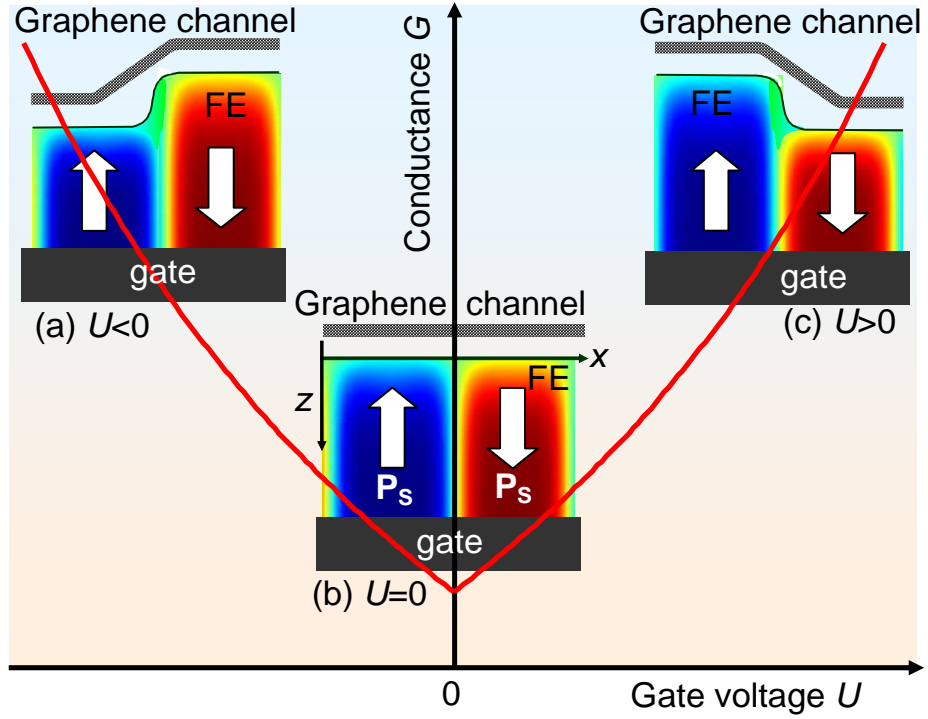

**Figure 5.4.** Modulation of graphene channel conductance by piezo-effect in GFeFET. Main plot shows the schematic dependence of the channel conductance $G(U)$ on the gate voltage $U$. Insets **(a)** and **(c)** illustrate the vertical piezoelectric displacement of ferroelectric surface at FDW that causes partial separation of graphene sections induced by negative and positive gate voltages, respectively. The displacement and corresponding graphene separation are absent at $U=0$ [Inset **(b)**]. Adapted from [A. N. Morozovska, A. I. Kurchak, .M. V. Strikha. Phys. Rev. Applied **8**, 054004 (2017)] with the permission of APS Publishing for the authors artwork.

Notably the separated sections of graphene channel induced by piezo-effect can cause the physical effects, which are interesting for fundamental physics. First, the conductance of graphene channel in diffusion regime changes significantly, because electrons in the stretched section scatter on acoustic phonons. Second, mechanic vibrations of MHz range can be realized here [73]. Third, high pseudo-magnetic fields were reported for stretched graphene [74].

## VI. CONCLUSIONS

Since GFETs on ferroelectric substrate are promising candidates for the novel non-volatile ultra-fast ferroelectric memories (FRAM) [, ], the theoretical description of the devices operation requires reliable knowledge about the ferroelectric response in a wide frequency range, as well as the nature of carrier trapping processes at a given frequency.

It has been shown that the contact between the domain wall and the ferroelectric surface creates a p-n junction in the graphene channel, at that that the carrier concentration induced in graphene by uncompensated ferroelectric dipoles can reach $10^{19}$ m$^{-2}$ in ballistic regime, that is two orders of magnitude higher than those obtained for the graphene on non-ferroelectric substrates [14]. The majority of realistic graphene devices are described by the mean free path of electrons ~ (50-250) nm and so are operating in a diffusive regime. Further we present the theory of the conductivity of p-n junctions in graphene channel, placed on ferroelectric substrate, caused by ferroelectric domain wall for the case of arbitrary current regime: from ballistic to diffusive one [15]. It had been demonstrated in [14, 15] that graphene channels with pnJ at FDW can serve as excellent rectifiers because of the great ratio of the conductance for the "direct" voltage applied to source and drain electrodes and the "opposite" one, caused by the large value of ferroelectric substrate permittivity.

The competition of the absorbed and ferroelectric dipoles would determine the FRAM operation characteristics. We propose a general theory for the analytical description of versatile hysteretic phenomena in a graphene field effect transistor (GFET) allowing for the existence of the external dipoles on graphene free surface and the localized states at the graphene-surface interface [16]. We demonstrated that the absorbed dipole molecules (e.g. dissociated or highly polarized water molecules) can cause hysteretic form of carrier concentration as a function of gate voltage and corresponding dependence of graphene conductivity in GFET on the substrate of different types, including the most common $SiO_2$ and ferroelectric ones. Results [16] are valid for the description of hysteretic phenomena in various realistic GFETs operating at low and intermediate frequency range, may be not directly applicable for the description of ferroelectric response at ultra-high frequencies, priory because the high-frequency response of absorbed and ferroelectric dipoles requires further experimental studies. The results obtained can be useful for prediction of the most suitable ferroelectric substrates for the graphene-on-ferroelectric based ultrafast non-volatile memory of new generation.

Using a self-consistent approach based on Landau-Ginzburg-Devonshire phenomenology combined with classical electrostatics we studied p-n junctions dynamics in graphene channel induced by stripe domains nucleation, motion and reversal in a ferroelectric substrate [17]. It was demonstrated, that for the case of intimate electric contact between the ferroelectric and graphene sheet relatively low gate voltages are required to induce the pronounced hysteresis of ferroelectric polarization and graphene charge in dependence on the periodic gate voltage. The electric boundary conditions for polarization at the lateral surfaces of ferroelectric substrate rules the asymmetry of the graphene channel conductance between the source and drain electrodes. Also we revealed the pronounced extrinsic size effect in the dependence of the graphene channel conductivity on its length.

P-n junctions in graphene on FDWs have been actively studied recently, but the role of piezoelectric effect in a ferroelectric substrate was not considered. We propose [18] a piezoelectric mechanism of conductance control in the GFET on a ferroelectric substrate with immobile domain walls. In particular we predict that the graphene channel conductance can be controlled by the gate voltage due to the piezoelectric elongation and contraction of ferroelectric domains with opposite polarization directions. At the same time the gate voltage create the bonded, separated, suspended and stretched sections of the graphene sheet, whose conductivity and resistivity are significantly different. Our calculations demonstrate the possibility of several times increase of GFET conductance for ferroelectric substrates with high piezoelectric response. Also we propose the alternative method of suspended graphene areas fabrication based on the piezo-effect in a ferroelectric substrate. The method does not require any additional technological procedures like chemical etching or mechanical treating of the substrate surface.

Taking into account that the conductance of the graphene-on-ferroelectric is significantly higher than the one of graphene on ordinary dielectric substrates, the predicted effect can be very useful for improvement and miniaturization of many types of electronic devices including various logic elements, memory cells, high efficient hybrid electrical modulators and voltage-to-current transducers with frequency doubling and relatively low operation voltages, and piezo-resistive elements.